\documentclass[pra,groupedaddress,showpacs,subeqn]{revtex4}
\usepackage{amsmath}  
\usepackage{graphicx}
\newcommand{\x}{{\bf x}}
\usepackage[dvips]{color}

\begin{document}

\title{
Highly nonlinear dynamics of third-harmonic generation
by focused beams}

\author{Richard S. Tasgal$^*$, Marek Trippenbach$^\dagger$,
M. Matuszewski$^\dagger$, and Y.B. Band$^*$}

\affiliation{
$^*$Departments of Chemistry and Electro-Optics,
Ben-Gurion University of the Negev, Beer-Sheva 84105, Israel \\
$^\dagger$Institute of Experimental Physics, Optics Division, Warsaw
University,
ul.~Ho\.{z}a 69, Warsaw 00-681, Poland}

\date{\today}

\begin{abstract}
Beams that experience third-harmonic generation (THG) also experience
Kerr effects.  With Kerr effects, beams do not take simple Gaussian
shapes, but exhibit nonlinear dynamics.  These nonlinear dynamics have
an effect on the THG accumulated by focusing and then diverging beams.
We formulate a self-consistent and complete set of nonlinear
Schr\"odinger equations for a pair of coupled beams -- a fundamental
and its third-harmonic.  Numerical simulations show that the Kerr
nonlinearities allow some third-harmonic to propagate to the far-field
even for zero or negative phase-mismatch.  This is because the
nonlinear dynamics break the beams' reflection symmetry about the focal
plane and therefore increases far-field THG by changing some of the
interference from destructive to constructive.  THG conversion
efficiencies are computed as functions of several beam parameters.

\end{abstract}



\maketitle

\section{Introduction}
We study the dynamics of two-color beams in a nonlinear isotropic
medium.  We take the initial conditions to be a one-color beam with
Gaussian profile.  With a cubic ($\chi^{(3)}$) nonlinearity, which is
the lowest order possible in an isotropic medium, the possible
nonlinearities are THG, Kerr, and Raman.  Any other nonlinearity
requires either a different susceptibility (second-harmonics,
high-harmonics), or more than two slowly-varying envelopes (sum and
difference frequency generation).  This is one of the simplest
nonlinear optics problems, and is for that reason important; it has an
application to third-harmonic generation (THG) microscopy
\cite{thg_microscopy}.

The usual model of THG takes the fundamental beam to be Gaussian, and
(inconsistently, or as an approximation) has THG as the sole
nonlinearity \cite{Ward.1969,Boyd.1992}.  This model has exact analytic
solution in which the third-harmonic (TH) beam takes a Gaussian
profile.  For a phase-mismatch that is zero or negative, the energy in
the TH peaks at the focus of the fundamental beam, and, after the
focus, destructive interference causes {\it all} the TH to be
re-absorbed by the fundamental.  The latter is often said to be due to
the Guoy shift \cite{Boyd.1992}, the phase-shift of $\pi$ that a
Gaussian beam experiences in going from a far-field, through a focus,
to a far-field.  But, in fact, the details show it to depend on the
beam's phase and shape everywhere along its path.  However, the fast
(electronic) nonlinear response of optical materials that yields THG
also generates Kerr effects (self- and cross-phase modulation).
Therefore, one should include Kerr and perhaps also Raman effects when
modeling THG. These can have quite drastic effects on a beam
\cite{Berge.1998}.  Although it was recognized as early as 1973 that
Kerr effects could influence THG \cite{Miles.1973}, and although Kerr,
THG, and dispersion have been examined in studies {\em without}
transverse spatial dynamics \cite{Band.1990,SammutBuryakKivshar.1998},
the results of transverse nonlinear dynamics on THG have not heretofore
been studied in quantitative detail.

We have derived a set of coupled nonlinear Schr\"odinger (NLS)
equations for two slowly-varying envelopes, with the relevant
nonlinearities treated rigorously and consistently
\cite{TasgalBand.nls_derivation}.  Because of the microscopy
application, we are interested in very tightly focused beams and
pulses, with large momentum and frequency spreads.  The NLS equations
are thus given to all orders in dispersion and diffraction; numerical
simulations are carried out with a method that is accurate to all
orders relevant for the grid
\cite{Trippenbach.2002,Rothenberg.1992,FleckMorrisFeit.1976}.  We
express the field as a fundamental, $A_{\omega}(\x,t)$, centered about
a carrier wave at frequency $\omega_0$, and a TH, $A_{3\omega}(\x,t)$,
centered about a carrier wave at frequency $3\omega_0$,
\begin{subequations}
\label{NLS:fund_and_th}
\begin{eqnarray}
    -i \partial_z A_{\omega}(\x,t)
    & = & \left[
           \sum_{p=1}^{\infty} \frac{i^p}{p!}
           \left(
             \frac{\partial^p}{\partial\omega^p} \beta(\omega)
           \right)_{\!\!\omega_0} \!\!\!
           \partial_t^p
         - \sum_{m=1}^{\infty} \frac{(2 m - 3)!!}{(-2)^{m} \, m!}
           \frac{\triangle_\perp^m}{\beta(\omega_0)^{2 m-1}}
          - \!\! \sum_{p,m=1}^{\infty}
           \frac{i^p \, (2 m - 3)!!}{(-2)^{m} \, m! \, p!}
           \left(
             \frac{\partial^p}{\partial\omega^p}
             \frac{1}{\beta(\omega)^{2 m-1}}
           \right)_{\!\!\omega_0} \!\!\!
           \triangle_\perp^m \partial_t^p
         \right]
         A_{\omega} \nonumber \\
    & & + \sum_{p,m=0}^{\infty}
         2\pi \frac{i^p \, (2 m - 1)!!}{(-2)^m \, m! \, p!}
         \left(
           \frac{\partial^p}{\partial\omega^p}
           \frac{(\omega/c)^2}{\beta(\omega)^{2 m+1}}
         \right)_{\omega_0} \!\!\!
         \triangle_\perp^m \partial_t^p \,
         P_{\omega} (\x,t) \; ,
         \label{NLS:fundamental} \\
    -i \partial_z A_{3\omega}(\x,t)
    & = & \left[
           \sum_{p=1}^{\infty} \frac{i^p}{p!}
           \left(
             \frac{\partial^p}{\partial\omega^p} \beta(\omega)
           \right)_{\!\!3\omega_0} \!\!\!\!\!
           \partial_t^p
         - \sum_{m=1}^{\infty} \frac{(2 m - 3)!!}{(-2)^{m} \, m!}
           \frac{\triangle_\perp^m}{\beta(3\omega_0)^{2 m-1}}
          - \!\! \sum_{p,m=1}^{\infty}
           \frac{i^p \, (2 m - 3)!!}{(-2)^{m} \, m! \, p!}
           \left(
             \frac{\partial^p}{\partial\omega^p} 
\frac{1}{\beta(\omega)^{2 m-1}}
           \right)_{\!\!3\omega_0} \!\!\!\!\!
           \triangle_\perp^m \partial_t^p
         \right]
         A_{3\omega} \nonumber \\
    & & + \sum_{p,m=0}^{\infty}
         2\pi \frac{i^p \, (2 m - 1)!!}{(-2)^m \, m! \, p!}
         \left(
          \frac{\partial^p}{\partial\omega^p}
          \frac{(\omega/c)^2}{\beta(\omega)^{2 m+1}}
         \right)_{\!\!3\omega_0} \!\!\!\!\!
         \triangle_\perp^m \partial_t^p \,
         P_{3\omega} (\x,t) \; .
         \label{NLS:th}
\end{eqnarray}
\end{subequations}
In the NLS equations~(\ref{NLS:fund_and_th}), $\triangle_\perp \equiv
\partial_x^2 + \partial_y^2$ is the transverse Laplacian, $n(\omega)$
is the index of refraction at frequency $\omega$, and $\beta(\omega)
\equiv n(\omega) \omega / c$ is the wavenumber.  The first summation
terms on the right-hand sides (RHS) of Eqs.~(\ref{NLS:fund_and_th}) are
dispersion, the second summation is diffraction, the third summation is
cross-dispersion/diffraction.  The nonlinear polarization includes THG,
Kerr, and Raman effects
\cite{SammutBuryakKivshar.1998,Boyd.1992,Agrawal.1995}.
Self-steepening terms \cite{steepening_refs,Trippenbach.1997}, of the
form $\partial (|A|^2 A)/\partial t$ and self-frequency shifting terms
of the form $\frac{\partial |A|^2}{\partial t} A$ are contained in the
first-order time derivatives of the nonlinear polarizations
$P_{\omega}(\x,t)$ and $P_{3\omega}(\x,t)$ in
Eqs.~(\ref{NLS:fund_and_th}).  The NLS equations
(\ref{NLS:fund_and_th}) show that self-steepening terms and
self-frequency shifting terms are just the first terms of a family of
higher order nonlinear terms.

The nonlinear polarization is taken to be of the form
\begin{subequations}
\label{nonlinear_polarization:Yehuda}
\begin{eqnarray}
    P_{\omega}(\x,t)
    & = & 3\chi^{\rm elec}(-\omega_0;-\omega_0,-\omega_0,3\omega_0) \;
          \exp\{ -i [3\beta(\omega_0) - \beta(3\omega_0)] z \} \;
          A_{\omega}(\x,t)^{* 2} A_{3\omega}(\x,t)
          \nonumber \\
    & + & 3\chi^{\rm elec}(-\omega_0;\omega_0,-\omega_0,\omega_0) \;
          |A_{\omega}(\x,t)|^2 A_{\omega}(\x,t)
        + 6\chi^{\rm elec}(-\omega_0;3\omega_0,-3\omega_0,\omega_0) \;
          |A_{3\omega}(\x,t)|^2 A_{\omega}(\x,t)
          \nonumber \\
    & + & A_{\omega}(\x,t)
          \int_0^\infty \!\!\!
          [ 3\chi^{\rm nucl}
            (-\omega_0;\omega_0,-\omega_0,\omega_0{\bf ;}\; s) \;
            |A_{\omega}(\x,t-s)|^2
          + 3\chi^{\rm nucl}
            (-\omega_0;\omega_0,-3\omega_0,3\omega_0{\bf ;}\; s)
            |A_{3\omega}(\x,t-s)|^2
          ] ds
          \nonumber \\
    & + & A_{3\omega}(\x,t)
          \int_0^\infty \!\!\! \exp(-2 i \omega_0 s)
          3\chi^{\rm nucl}
          (-\omega_0;3\omega_0,-\omega_0,-3\omega_0{\bf ;}\; s)
          A_{\omega}(\x,t-s) A_{3\omega}(\x,t-s)^*
          ds  \\
    P_{3\omega}(\x,t)
    & = & \chi^{\rm elec}(-3\omega_0;\omega_0,\omega_0,\omega_0) \;
          \exp\{i [3\beta(\omega_0) - \beta(3\omega_0)] z \} \;
          A_{\omega}(\x,t)^3
          \nonumber \\
    & + & 6\chi^{\rm elec}(-3\omega_0;\omega_0,-\omega_0,3\omega_0) \;
          |A_{\omega}(\x,t)|^2 A_{3\omega}(\x,t)
        + 3\chi^{\rm elec}(-3\omega_0;3\omega_0,-3\omega_0,3\omega_0) \;
          |A_{3\omega}(\x,t)|^2 A_{3\omega}(\x,t)
          \nonumber \\
    & + & A_{3\omega}(\x,t)
          \int_0^\infty \!\!\!
          [ 3\chi^{\rm nucl}
            (-3\omega_0;\omega_0,-\omega_0,3\omega_0{\bf ;}\; s) \;
            |A_{\omega}(\x,t-s)|^2
          + 3\chi^{\rm nucl}
            (-3\omega_0;3\omega_0,-3\omega_0,3\omega_0{\bf ;}\; s) \;
            |A_{3\omega}(\x,t-s)|^2
          ] ds
          \nonumber \\
    & + & A_{\omega}(\x,t)
          \int_0^\infty \!\!\!  \exp(2 i \omega_0 s) \,
          3\chi^{\rm nucl}
          (-3\omega_0;\omega_0,-\omega_0,3\omega_0{\bf ;}\; s) \;
          A_{\omega}(\x,t-s)^* A_{3\omega}(\x,t-s)
          ds \; .
\end{eqnarray}
\end{subequations}
This breaks up the nonlinear response into an electronic (fast) part
and a nuclear (slow) part.  It also assumes that the electronic part of
the response may be considered frequency-independent on the scale of
the pulse bandwidths; it does not make this assumption for the nuclear
part of the response.  For the calculations, we take a more specific
nonlinear polarization, essentially the standard model of the
third-order susceptibility of fused silica
\cite{Hellwarth.1977,Stolen.1989,Blow.1989,Agrawal.1995} plus a
generalization,
\begin{subequations}
\label{nonlinear_polarization:Kerr_Raman_thg}
\begin{eqnarray}
   P_{\omega}(\x,t)
   & = & \exp\{ -i [3\beta(\omega_0) - \beta(3\omega_0)] z \} \;
         \chi^{\rm THG}
         A_{\omega}(\x,t)^{* 2} A_{3\omega}(\x,t)
       + \chi^{\rm elec} 
         (|A_{\omega}(\x,t)|^2 + 2 |A_{3\omega}(\x,t)|^2) A_{\omega}(\x,t)
       \nonumber \\
   & & + A_{\omega}(\x,t) \int_0^\infty \!\!\! \chi^{\rm nucl}(s)
        (|A_{\omega}(\x,t-s)|^2 + |A_{3\omega}(\x,t-s)|^2) ds
       \nonumber \\
   & & + A_{3\omega}(\x,t) \int_0^\infty \!\!\! \exp(-2 i \omega_0 s) \chi^{\rm nucl}(s)
       A_{\omega}(\x,t-s) A_{3\omega}(\x,t-s)^* ds \,,  \\
   P_{3\omega}(\x,t)
   & = & \exp\{i [3\beta(\omega_0) - \beta(3\omega_0)] z \} \;
         \frac{1}{3} \chi^{\rm THG} A_{\omega}(\x,t)^3
       + \chi^{\rm elec}
         (2 |A_{\omega}(\x,t)|^2 + |A_{3\omega}(\x,t)|^2) A_{3\omega}(\x,t)
       \nonumber \\
   & & + A_{3\omega}(\x,t) \int_0^\infty \!\!\! \chi^{\rm nucl}(s)
         (|A_{\omega}(\x,t-s)|^2 + |A_{3\omega}(\x,t-s)|^2) ds
       \nonumber \\
   & & + A_{\omega}(\x,t) \int_0^\infty \!\!\! \exp(2 i \omega_0 s) \chi^{\rm nucl}(s)
         A_{\omega}(\x,t-s)^* A_{3\omega}(\x,t-s) ds \, .
\end{eqnarray}
\end{subequations}
Here the THG coefficient is decoupled from the other electronic
susceptibilities; this is outside the usual model for pure fused silica
in which the electronic contribution is considered instantaneous and
the nuclear contrubution takes the form arising from a single damped
harmonic oscillator, $\chi(t;t_1,t_2,t_3) = \chi^{\rm elec}\,
\delta(t-t_1) \delta(t_1 - t_2) \delta(t_2 - t_3) + \chi^{\rm
nucl}(t_1-t_2) \delta(t - t_1) \delta(t_2 - t_3)$, with $\chi^{\rm
elec} = n_2 (1 - f_{\rm Raman})$, $\chi^{\rm nucl}(t) = n_2 f_{\rm
Raman} (\tau_1^2+\tau_2^2) \tau_1^{-1} \tau_2^{-2} \exp(-t/\tau_2)
\sin(t/\tau_1)$, $n_2 = n(\omega) c\, n_2^I / (2\pi)$, $n_2^I = 2.8
\times 10^{-20}$ m$^2$/W, $f_{\rm Raman} = 0.18$, $\tau_1 = 12.2$ fs,
and $\tau_2 = 32.0$ fs.  If the electronic THG susceptibility differs
from the other electronic susceptibilities, the electronic contribution
to the susceptibility is not instantaneous compared to all scales.
Direct experimental measurements of the THG susceptibilities are
available \cite{Milam.1998,GublerBosshard.2000,THG.older_measurements}.
The more recent measurement of the THG coefficient $\chi^{\rm THG}$
\cite{GublerBosshard.2000} is smaller than the electronic contribution
to self-phase modulation \cite{Milam.1998} $\chi^{\rm elec}$ by a
factor of almost four; older measurements \cite{THG.older_measurements}
give a THG coefficient smaller than the electronic part of the
self-phase modulation by a factor of about $1.5$.  In the absense of
direct experimental measurements of all the nonlinear polarization
coefficients for doped silica, we use the simplest case $\chi^{\rm THG}
= \chi^{\rm elec}$ in numerical simulations, but discuss how the
results scale for different values of the THG coefficient.
%
%
%
%
Vector effects are neglected.  Inter-band Raman scattering is
negligible {\it at the carrier frequencies} because of the fast
relative phase oscillation; but in the pulse simulations, inter-band
Raman scattering is possible between the lower frequencies within the
higher frequency band, and the higher frequencies within the lower
frequency band.
%

\section{Numerical Simulations}
A numerical NLS propagation scheme may be said to be accurate to all
orders of dispersion and diffraction if it is accurate to as many
orders as there are grid points.  Accuracy up to all available orders
requires the index of refraction over the entire numerically
represented frequency range.  In the split-step fast Fourier transform
scheme, linear propagation is carried out in momentum space; the
algorithm may be made accurate to all orders by putting the frequency
dependent index of refraction directly into the formulas for the
propagators \cite{TasgalBand.nls_derivation}, i.e.,
\begin{subequations}
\begin{eqnarray}
    -i \partial_z A_\omega(\x,t)
    & = & {\cal F}^{-1}
            \left\{
              \left[
                \sqrt{\beta(\omega_0+\omega)^2 - k_\perp^2}
              - \beta(\omega_0)
              \right] \, {\cal F} \{ A_\omega(\x,t) \}
            \right\}
        + {\cal F}^{-1}
             \left\{
               \frac{2\pi \, (\omega_0+\omega)^2 / c^2}
                    {\sqrt{\beta(\omega_0+\omega)^2 - k_\perp^2}}
               {\cal F} \{ P_\omega (\x,t) \}
             \right\} \; , \\
    -i \partial_z A_{3\omega}(\x,t)
    & = & {\cal F}^{-1}\!\!
            \left\{
              \left[
                \sqrt{\beta(3\omega_0+\omega)^2 - k_\perp^2}
              - \beta(3\omega_0)
              \right] {\cal F} \{ A_{3\omega}(\x,t) \}
            \right\}
        + {\cal F}^{-1}\!\!
            \left\{
              \frac{2\pi \, (3\omega_0+\omega)^2 / c^2}
                   {\sqrt{\beta(3\omega_0+\omega)^2 - k_\perp^2}}
              {\cal F} \{ P_{3\omega} (\x,t) \}
            \right\} \, .
\end{eqnarray}
\end{subequations}
Here ${\cal F}$ and ${\cal F}^{-1}$ are Fourier and inverse Fourier
transforms in $x,y,t$, $k_\perp = \sqrt{k_x^2+k_y^2}$ is the transverse
momentum, and $\omega$ is frequency.  The linear dispersion effects are
contained in $\beta(\omega) = n(\omega) \omega / c$; diffraction is
contained in $k_\perp$.  The offsets by $\omega_0$ and $3\omega_0$ are
due to the fact that the fields are slowly-varying envelopes about
carrier waves at those frequencies; when these appear, $\omega$ is the
frequency relative to the offset.  There is no need to compute the
coefficients for dispersion, diffraction, self-steepening, {\it etc}.,
explicitly because they are contained implicitly in the linear
dispersion $\beta(\omega)$.  Since we simulate focusing and collapse,
during which diffractive, dispersive, and nonlinear length scales can
easily change by factors of a hundred or more
\cite{Trippenbach.1997,Agrawal.1995}, we allow the propagation step to
vary such that it remains an order of magnitude less that the smallest
relevent scale.

The effects that we are interested in are larger and clearer when the
fundamental and TH are not too far from being phase-matched, $\Delta k
\equiv 3\beta(\omega_0) - \beta(3\omega_0) \approx 0$, or $n(\omega_0)
\approx n(3\omega_0)$.  Optical materials may be doped to obtain
desired properties \cite{Agrawal.1995,Nicacio.1993}.  We consider
silica doped with neodynium to obtain approximate phase-matching
between the fundamental $\lambda=1.5\, \mu$m and TH $\lambda=0.5\,
\mu$m.  To model the frequency dependent index of refraction, we use
the Sellmeier relation for fused silica \cite{Malitson.1965}, but add
one additional resonance at $\lambda_{\rm Nd} = 0.59\, \mu$m, which is
neodynium's largest resonance in the vicinity of our TH. For this,
phase-matching is achieved at the Sellmeier coefficient $B_{\rm Nd}
\approx 0.0138$, which corresponds to a few percent doping of the
material.  Phase-mismatch is varied by changing the dopant
concentration (in this model, the Sellmeier coefficient $B_{\rm Nd}$).
We take the conventional nonlinear coefficients for pure silica
\cite{Agrawal.1995}, as given above.
%
%

\begin{figure}
\centerline{\scalebox{1}{\includegraphics{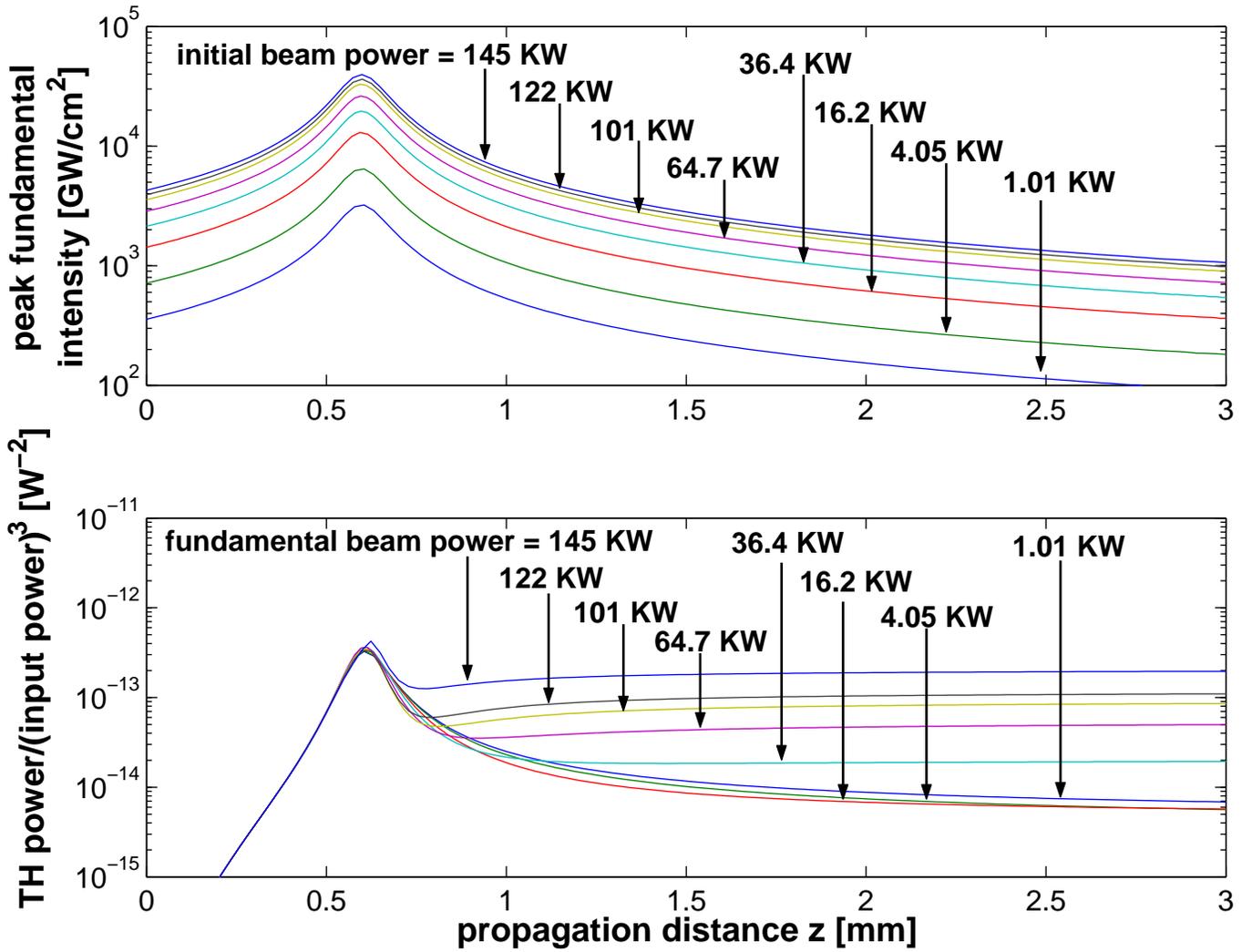}}}
\caption{Peak intensity of the fundamental ($\lambda_1 = 1.5 \, \mu$m)
beam, and power of the third-harmonic ($\lambda_3 = 0.5 \, \mu$m)
normalized by the cube of the input power.  The initial conditions have
a range of intensities (1.01, 4.05, 16.2, 36.4, 64.7, 101, 122 and 145
KW, corresponding to circled points in the appropriate curve in
Fig.~\ref{fig2}), but identical beam width, FWHM = $50\, \mu$m, radial
phase factor, $c_1 = 5.0 \times 10^9 / {\rm m}^2$ (i.e., transverse
phase $\exp[-i c_1 (x^2+y^2)]$), and all are phase-matched.}
\label{fig1}
\end{figure}

We simulated the propagation of both pulses and continuous beams, over
a range of light intensities, phase-mismatch values, and focusing
strengths.  The effects of the nonlinearity are clearly visible in
Fig.~\ref{fig1}, which shows the evolution of beams with varying
intensity.  For all simulations, we took initial conditions with zero
power in the TH, and the fundamental in the form of a Gaussian with
some radial phase factor $A_1(x,t;z=0) = A_1 \exp[-(i c_1 + 2\ln
2/W_1^2)(x^2+y^2)]$.  The radial phase factor $c_1$ can be related to
the wavefront radius of curvature $R_1$ via the magnitude of the
wavevector $\beta(\omega) = n(\omega) \omega/c$ according to the
formula $R_1 = 0.5 \beta(\omega) / c_1$.  Because the fundamental
focuses very intensely, while only a relatively small part of it is
converted to TH, we illustrate the {\it peak intensity} of the
fundamental beam and the {\it power} of the TH beam.  In
Fig.~\ref{fig1}, the input beam has initial width FWHM = $50 \, \mu$m,
a radial phase factor $c_1 = 5.0 \times 10^9 / {\rm m}^2$ that brings
the beam to a focus in 0.6 mm, and the material is phase-matched (the
reciprocal of the phase-mismatch is more than an order of magnitude
greater than the simulation distance).  The fact that the {\em
normalized} TH power varies with fundamental input intensity shows that
the beams experience nonlinear dynamics, changing shape as the
intensity changes.  In Fig.~\ref{fig1}, the initial wave front radii of
curvature are small and their effect on the position of the focus
overshadow the intensity dependence.  One can see that the beams
approach the linear limit at low intensities, as these curves start to
almost overlap.  Fig.~\ref{fig2} shows the normalized {\it far-field}
TH power as a function of input power, for the data in Fig.~\ref{fig1}
and for another set of runs with weaker focusing.

For a qualitative explanation of these figures, we first contrast our
results with the model in which THG is the only nonlinearity
\cite{Ward.1969,Boyd.1992}.  Here, in the case with phase-matching,
the TH power $P_{\rm TH}(z) = (3/2) (2\pi/n c)^2 (\omega/c)^4 (\pi
W_1^2)^3 {\chi^{(3)}}^2 I_{\omega}^3 / [1 + (z-z_{\rm
focus})^2/z_1^2]$ reaches a maximum at the focus of the fundamental
and then drops off as the inverse square of $z/z_1$, where $z_1 = \pi
W_1^2 / \lambda_0$ is the Rayleigh range of the fundamental {\it and}
the TH beam, and $W_1$ is the width of the fundamental at its focus.
Our simulations show that at high intensities, the Kerr effect causes
the beams to lose reflection symmetry about the focal plane.  With
this asymmetry, THG from the incoming and outgoing beams does not
fully interfere destructively, and allows some TH to propagate to the
far-field.  The lower-intensity curves in Fig.~\ref{fig1} do not quite
drop off as the inverse square of distance because the numerical
simulations did not start out from minus infinity, but began with
merely a large (finite) beam width.  The reflection asymmetry in
Fig.~\ref{fig1} is due partly to the nonlinearities and partly to
starting with a finite initial beam width.  As long as the amount of
energy in the TH is relatively small, the nonlinear dynamic effects
will remain in the fundamental beam, and the TH peak intensities and
beam powers may scale up or down by a uniform factor, but will be
otherwise unaffected.  Clearly, there is significant variation with 
the fundamental input power.  Moreover, this variation is quite 
different for different focusing conditions.

In another series of simulations, phase-mismatch is varied.
Fig.~\ref{fig3} shows the peak intensity of the fundamental and the TH
beam power with initial FWHM = $50\, \mu$m, radial phase-factor $c_1 =
5.0 \times 10^9/{\rm m}^2$, and phase-mismatch from -12.9 to 51.8
mm$^{-1}$.  Because the TH power first increases and then decreases
with phase-mismatch, we show the former range on one plot and the
latter on another.  Clearly, there is a residual far-field TH and its
power depends on phase-mismatch.  Fig.~\ref{fig4} shows the far-field
TH conversion efficiency as a function of phase mismatch, for the runs
in Fig.~\ref{fig3} and another set of runs with weaker focusing, $c_1 =
2.5 \times 10^9/{\rm m}^2$.  As in the linear model \cite{Ward.1969},
THG by a focusing beam is maximized around a certain phase-mismatch;
but the nonlinear dynamics complicate the results considerably.

In a further series of simulations, the intensity and phase-mismatch
were held constant, and the radial phase factor varied.  This mixes up
a few physical effects, since initial conditions with small radial
phase factors cannot be considered as starting from the far-field.
Fig.~\ref{fig5} has extremely small phase-mismatch ($\Delta k = 0.067$
mm$^{-1}$), and radial phase factors which vary from zero (i.e.,
starting at a focus) up to $c_1 = 5.0 \times 10^9/{\rm m}^2$.  The
horizontal axis is on a log scale to help visually distinguish the
superimposed sumulations.  The peak fundamental intensity varies, but
over the range studied, the peak TH power is rather insensitive to the
wavefront radius of curvature of the fundamental, but the far-field TH
power does vary considerably.  Fig.~\ref{fig6} shows the far-field TH
power as a function of radial phase factor, for the runs in
Fig.~\ref{fig5}, and also for a series of runs with significant
positive and negative phase-mismatch.  In increasing (decreasing) the
tightness of the focusing, the nonlinear and diffraction lengths both
decrease (increase); the complexity of the results for far-field THG
reflect the complexity of the nonlinear dynamics of the beam.

Fig.~\ref{fig7} shows the dependence of the third-harmonic power on the
when the THG suseptibility $\chi^{\rm THG}$ is varied.  The curve
labeled $\chi^{\rm THG} = \chi^{\rm elec}$ corresponds to the same
conditions used in Fig.~\ref{fig3} but with $\Delta k = 0.067$
mm$^{-1}$ (i.e., almost phase-matched).  The third harmonic power
scales with $|\chi^{\rm THG}|^2$ when the third-harmonic intensity is
small since then the nonlinear dynamics of the fundamental is
unaffected by TH and the generation of TH field is proportional to
$\chi^{\rm THG}$.  The values of $\chi^{\rm THG}$ used in the
calculations shown in Fig.~\ref{fig7} are a factor of $1.0$, $1/3.9$,
and $1/1.5$ times the value used in the previous figures, where the
latter two factors correspond to the measured values of $\chi^{\rm
THG}$ reported in
Refs.~\cite{Milam.1998,GublerBosshard.2000,THG.older_measurements}

Numerical simulations showed the dynamics of pulses to differ from
those of continuous waves in essentially two ways.  First, the group
velocities of the fundamental and TH will, except for special cases,
not be the same.  A TH pulse will thus generally walk off from a
fundamental pulse.  This tends to reduce destructive interference; it
also limits the effective distance over which the pulses interact.
Compared to beams, TH pulses tend to carry off a larger part of the
fundamental pulse energy.  Secondly, the dynamics of pulses
(3+1--dimensional) are qualitatively different to those of continuous
beams (2+1--dimensional).  For strongly focused but not very short
pulses, these differences tend to be minor.  A thorough analysis of
focused pulses in this system is quite involved, and will be pursued
elsewhere.  Thus, the detailed results for continuous-wave beams apply
to pulses for which group-velocity differences are relatively small or
for pulses that are relatively long.

\section{Conclusions}
In a nonlinear medium, intense beams or pulses of finite diameter which
converge to a focus and then diverge may exhibit nonlinear dynamics
that significanly affect propagation dynamics.  These nonlinear effects
break the beam's reflection symmetry about the focal plane.  The
greater the intensity, the bigger the difference between incoming and
outgoing beams.  THG with such an input beam or pulse produces TH in
the far-field when the phase-mismatch between the fundamental and its
third-harmonic is zero or negative (and small).  We have quantitatively
demonstrated this for several cases.  When phase-mismatch is positive,
where some far-field TH power is possible in the essentially linear
case, the nonlinear beam dynamics complicate the accumulation of TH
power.  The non-zero far-field THG for zero or negative phase-mismatch
-- a qualitatively new effect for homogeneous media -- is affected by,
and thus contains information about the medium in the region of the
beam focus.  For an inhomogeneous medium, and for THG microscopy, these
effects should be understood, either to be utilized or better avoided.

\begin{acknowledgments}
Richard Tasgal gratefully acknowledges a Kreitman Foundation
Fellowship.  This work was supported in part by a grant from the
Israel Science Foundation for a Center of Excellence (grant No.
8006/03) and by KBN as a reasearch grant 2003-2006 (2P03B04325).
\end{acknowledgments}

\newpage

\section*{List of Figure Captions}

\begin{figure}
\centerline{\scalebox{1}{\includegraphics{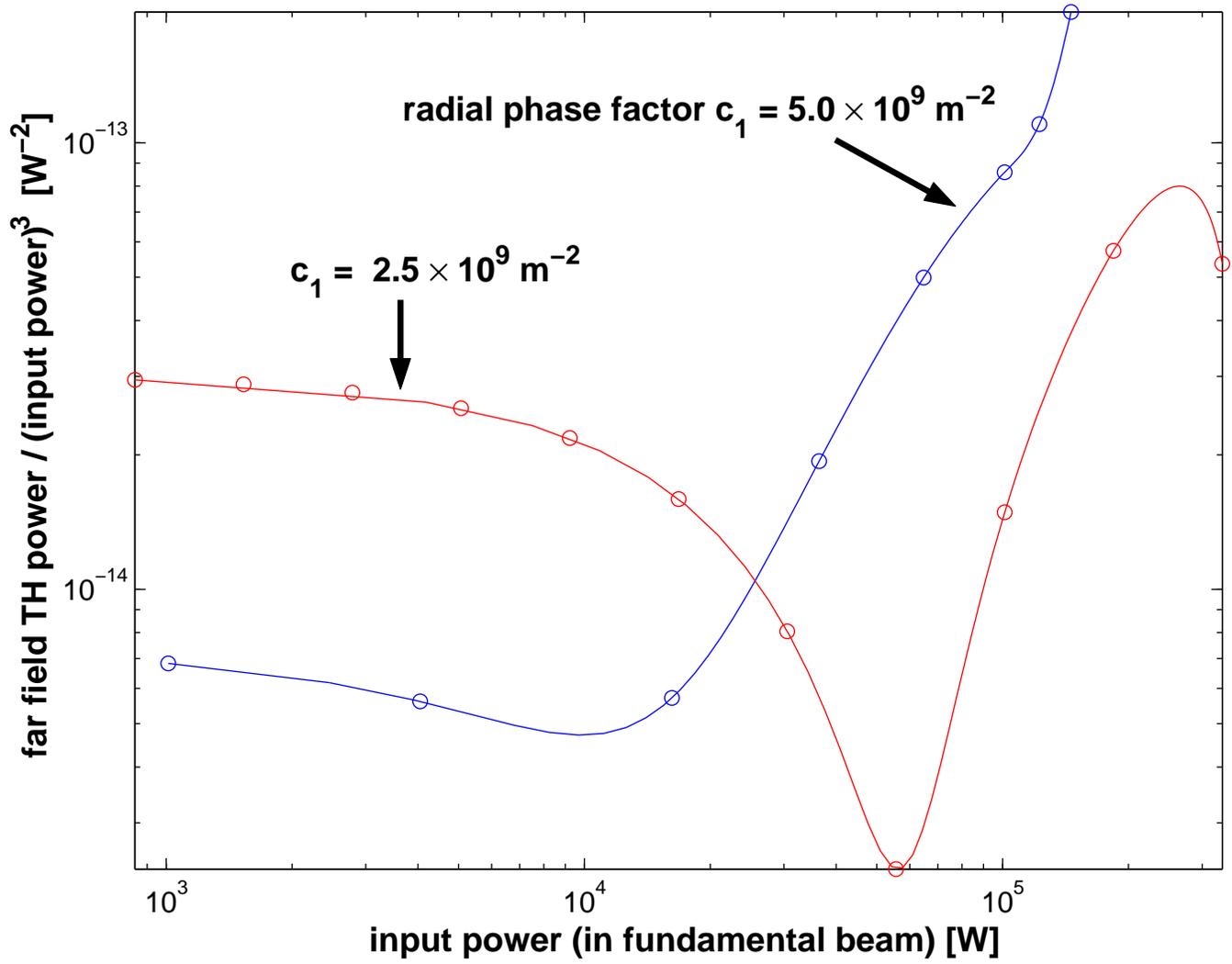}}}
\caption{Far-field third-harmonic beam power, normalized by the cube
of the input power, {\it vs}.\ input power.  The medium has TH
phase-matching.  Two curves are shown, one the far-field results from
Fig.~\ref{fig1}, and another with smaller radial phase factor, that
brings the beam to a focus in about 1.2 mm.}
\label{fig2}
\end{figure}

\begin{figure}
\centerline{\scalebox{1}{\includegraphics{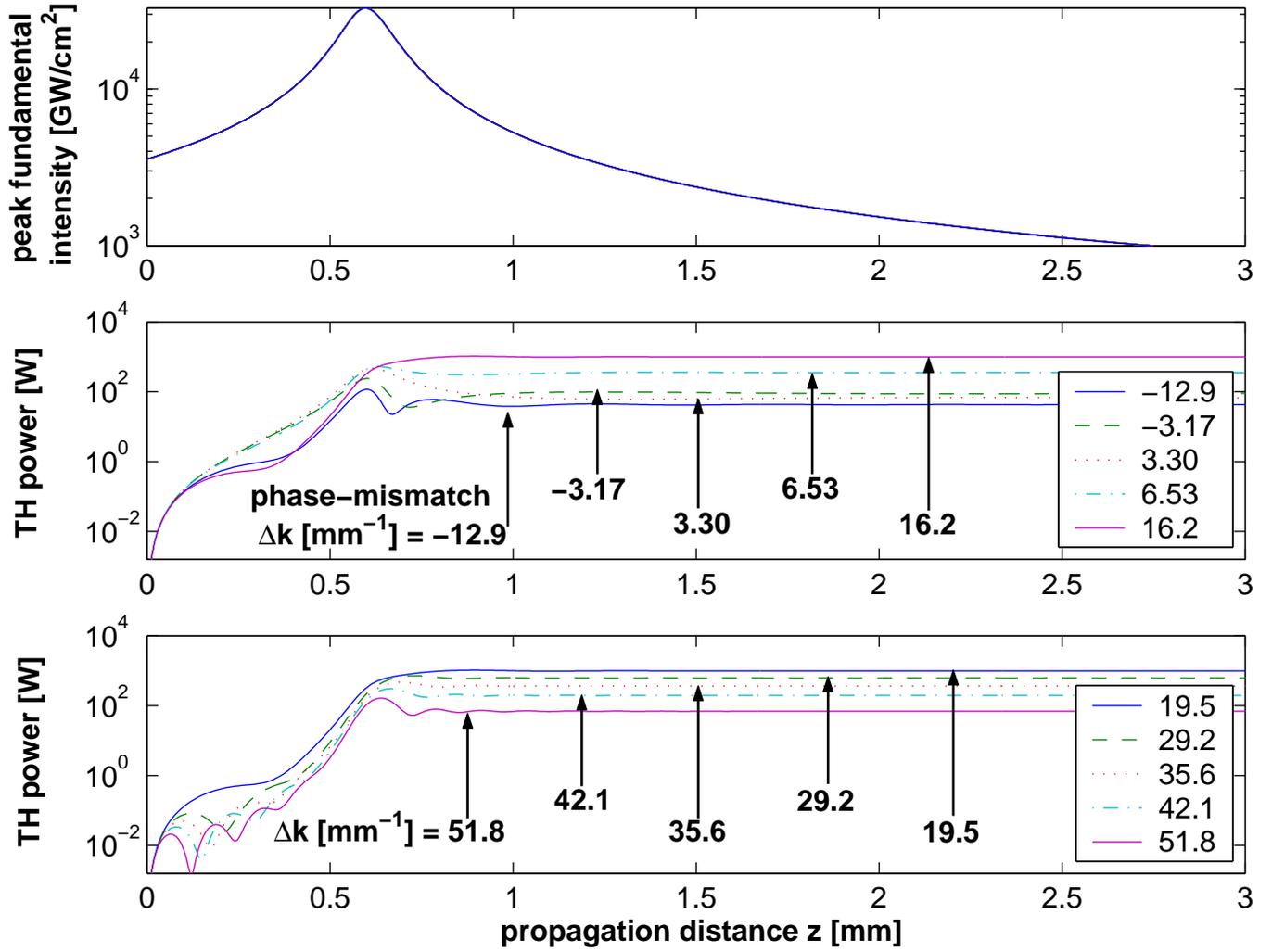}}}
\caption{Peak intensity of the fundamental beam, and power of the
third-harmonic.  The curves represent beams with identical power,
width, and radial phase factor, while the phase-mismatch goes from
-12.9 to 51.8 mm$^{-1}$ (the specific values correspond to circled
points in the appropriate curve in Fig.~\ref{fig4}).  The middle plot
shows the TH increasing with phase-mismatch, and the bottom plot shows
TH decreasing with phase-mismatch.}
\label{fig3}
\end{figure}

\noindent Fig.~1.
Peak intensity of the fundamental ($\lambda_1 = 1.5 \, \mu$m) beam,
and power of the third-harmonic ($\lambda_3 = 0.5 \, \mu$m) normalized
by the cube of the input power.  The initial conditions have a range
of intensities (1.01, 4.05, 16.2, 36.4, 64.7, 101, 122 and 145 KW,
corresponding to circled points in the appropriate curve in
Fig.~\ref{fig2}), but identical beam width, FWHM = $50\, \mu$m, radial
phase factor, $c_1 = 5.0 \times 10^9 / {\rm m}^2$ (i.e.,
transverse phase $\exp[-i c_1 (x^2+y^2)]$), and all are
phase-matched. \\

\noindent Fig.~2.
far-field third-harmonic beam power, normalized by the cube of the
input power, {\it vs}.\ input power.  The medium has TH
phase-matching.  Two curves are shown, one the far-field results from
Fig.~1, and another with smaller radial phase factor, that brings the
beam to a focus in about 1.2 mm. \\

\begin{figure}
\centerline{\scalebox{1}{\includegraphics{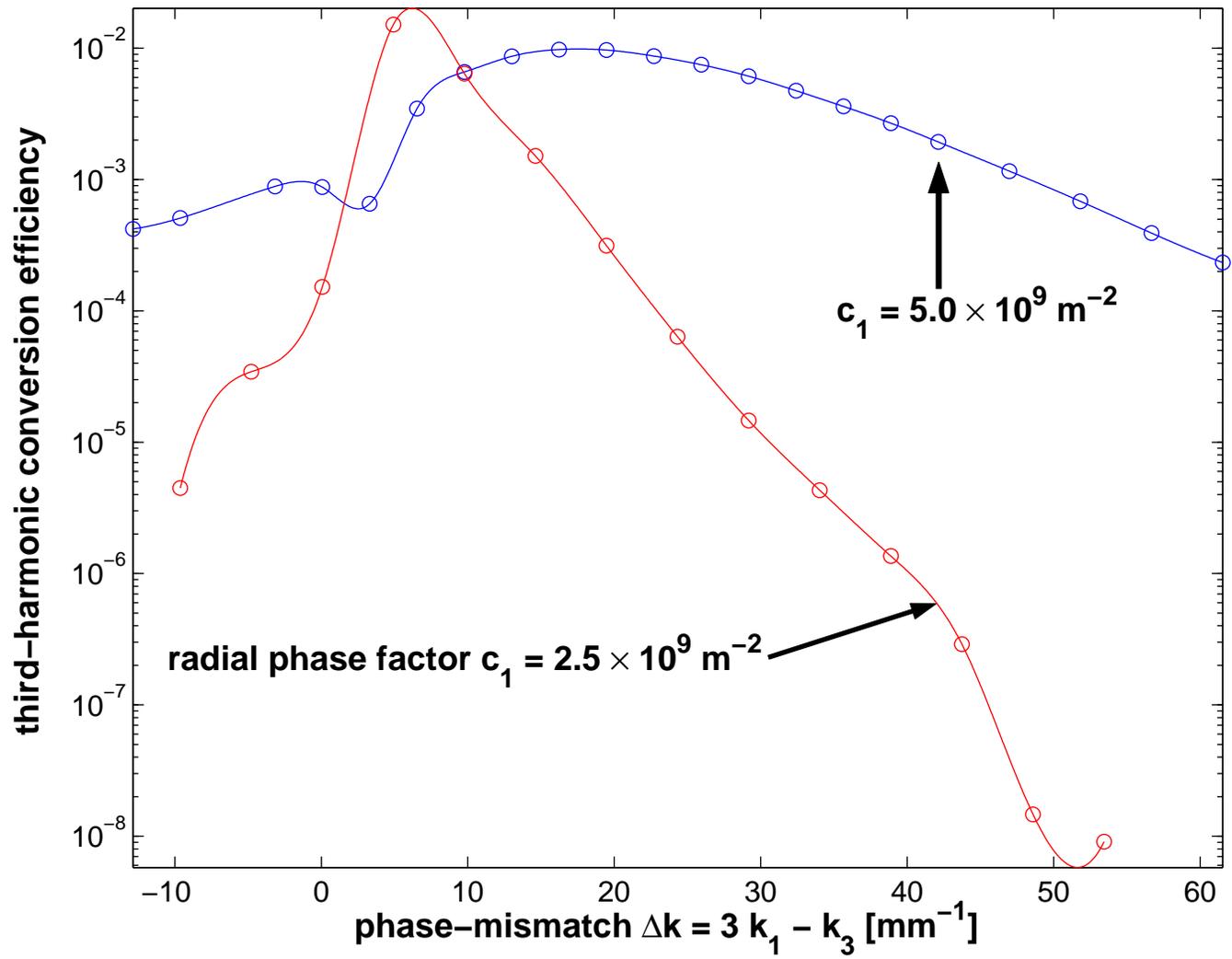}}}
\caption{Far-field conversion efficiency {\it vs}.\ phase-mismatch.
Two curves are shown, one the far-field results from Fig.~\ref{fig3},
and another with smaller radial phase factor, that brings the beam to
a focus in about 1.2 mm.  Initial intensities are held constant.}
\label{fig4}
\end{figure}

\begin{figure}
\centerline{\scalebox{1}{\includegraphics{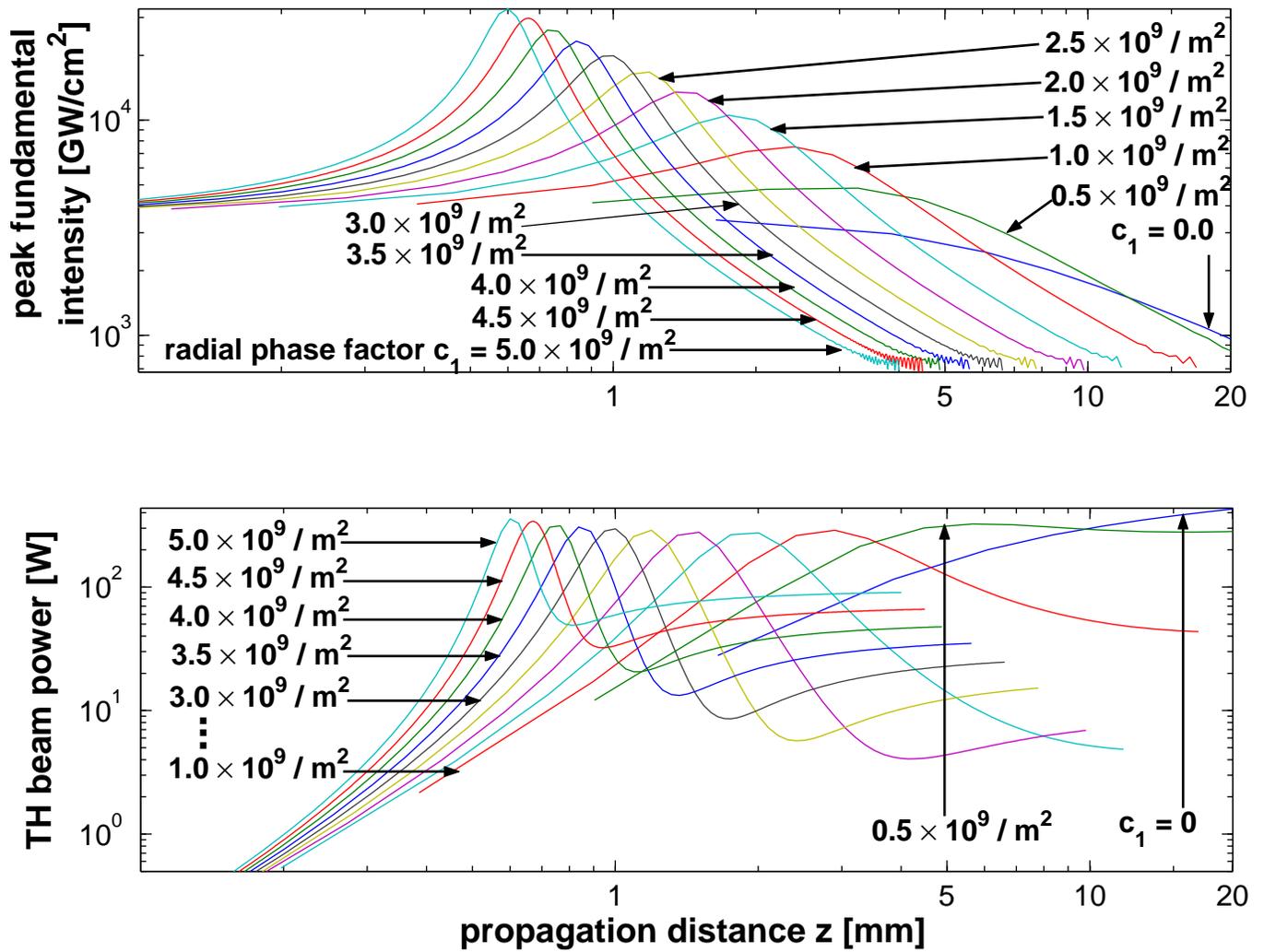}}}
\caption{Peak intensity of the fundamental beam, and power of the
third-harmonic versus propagation distance for different fundamental
radial curvatures.  The curves show beams with phase-matching and
initially identical power and width; the initial radial phase factors
go from zero (starting at a focus) up to $5.0 \times 10^9/{\rm m}^2$,
by increments of $0.5 \times 10^9/{\rm m}^2$ (the curves here
correspond to circled points in the appropriate curve in
Fig.~\ref{fig6}).}
\label{fig5}
\end{figure}

\noindent Fig.~3 Peak intensity of the fundamental beam, and power of
the third-harmonic.  The curves represent beams with identical power,
width, and radial phase factor, while the phase-mismatch goes from
-12.9 to 51.8 mm$^{-1}$
%
%
%
%
%
(corresponding to circled points in the appropriate curve in
Fig.~\ref{fig4}).  The middle plot shows the TH increasing with
phase-mismatch, and the bottom plot shows TH decreasing with
phase-mismatch.  \\

\noindent Fig.~4.
Far-field conversion efficiency {\it vs}.\ phase-mismatch.  Two curves
are shown, one the far-field results from Fig.~3, and another with
smaller radial phase factor, that brings the beam to a focus in about
1.2 mm.  Initial intensities are held constant.  \\

\begin{figure}
\centerline{\scalebox{1}{\includegraphics{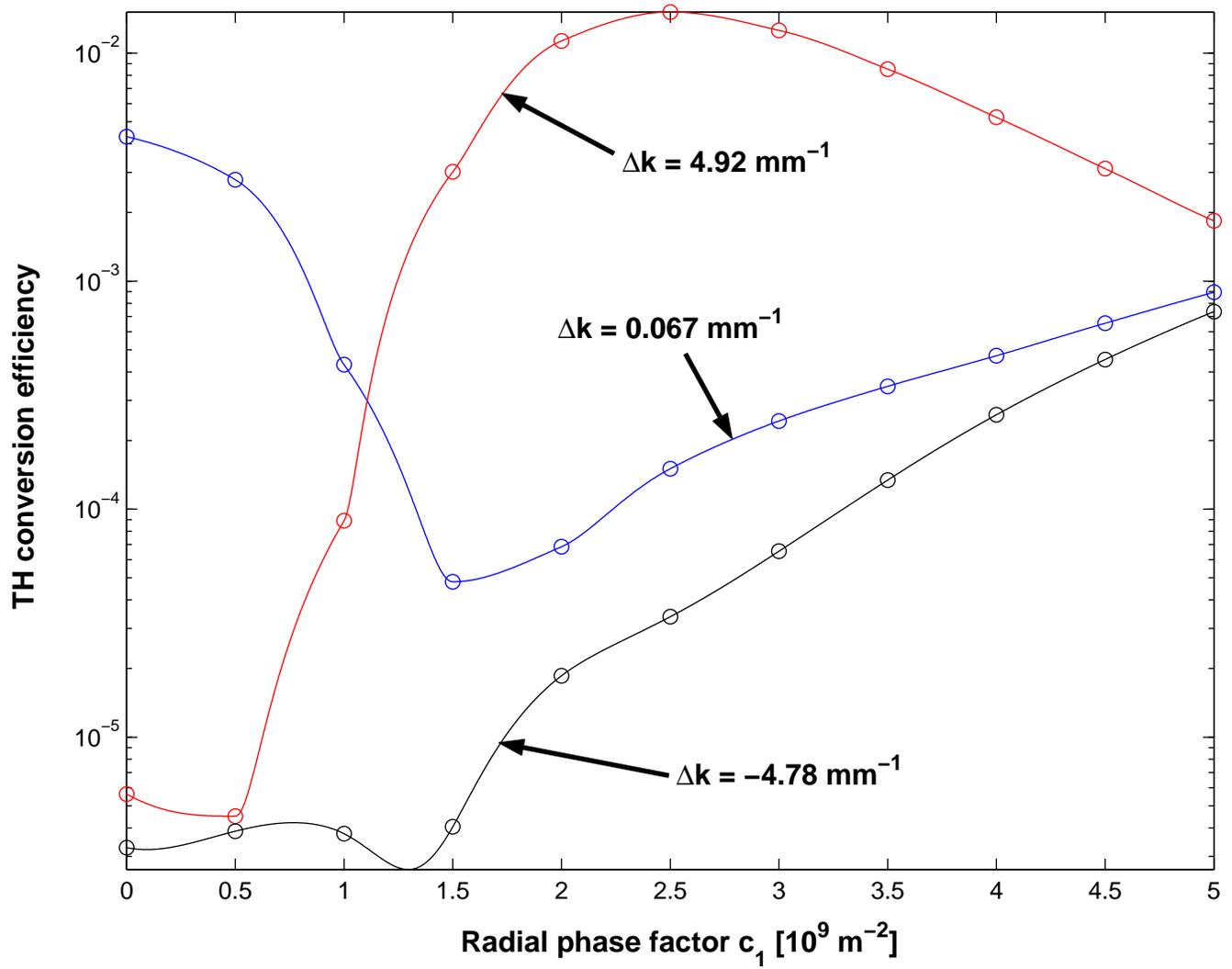}}}
\caption{Far-field conversion efficiency {\it vs}.\ radial phase
factor.  Three curves are shown, one corresponding to the far-field
results of Fig.~\ref{fig5} (phase-matching), and two others
corresponding to positive and negative phase-mismatch.}
\label{fig6}
\end{figure}

\begin{figure}
\centerline{\scalebox{1}{\includegraphics{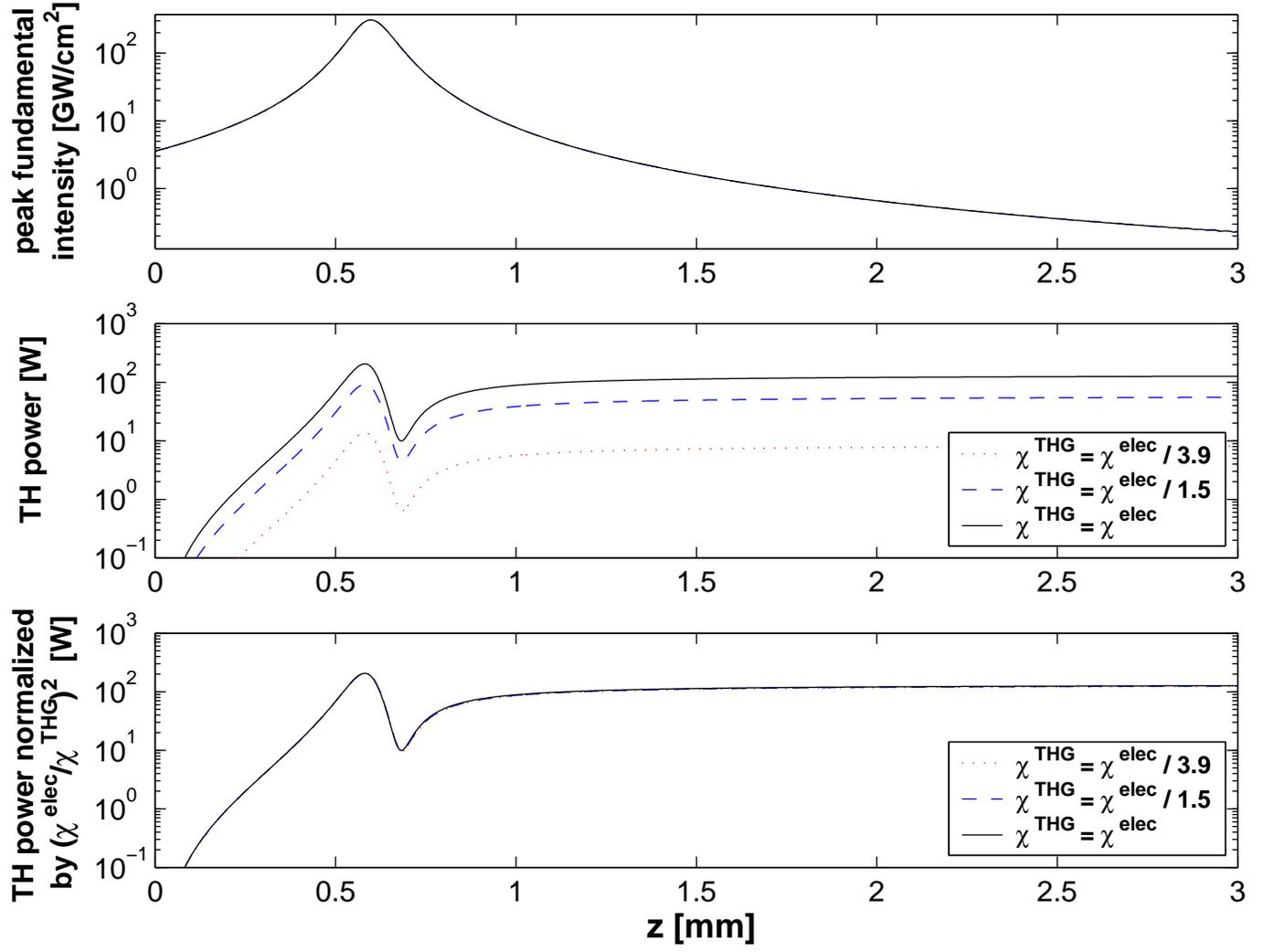}}}
\caption{Peak fundamental intensity and third-harmonic power versus
position $z$, for various values of $\chi^{\rm THG}$.  The third
harmonic power scales with $|\chi^{\rm THG}|^2$ when the third-harmonic
intensity is small.}
\label{fig7}
\end{figure}

\noindent Fig.~5.  Peak intensity of the fundamental beam, and power of
the third-harmonic versus propagation distance for different
fundamental radial curvatures.  The curves show beams with
phase-matching and initially identical power and width; the initial
radial phase factors go from zero (starting at a focus) up to $5.0
\times 10^9/{\rm m}^2$, by increments of $0.5 \times 10^9/{\rm m}^2$
(the curves here correspond to circled points in the appropriate curve
in Fig.~\ref{fig6}).  \\

\noindent Fig.~6.
Far-field conversion efficiency {\it vs}.\ radial phase factor.
Two curves are shown, one the far-field results from Fig.~5
(phase-matching), and another in a medium with positive
phase-mismatching.

\noindent Fig.~7.  Peak fundamental intensity and third-harmonic power
versus position $z$, for various values of $\chi^{\rm THG}$.  The third
harmonic power scales with $|\chi^{\rm THG}|^2$ when the third-harmonic
intensity is small.


\end{document}